\documentclass[sigconf,anonymous=false]{acmart}
\usepackage{booktabs}
\usepackage{multirow}

\AtBeginDocument{%
  \providecommand\BibTeX{{%
    \normalfont B\kern-0.5em{\scshape i\kern-0.25em b}\kern-0.8em\TeX}}}

\copyrightyear{2024} 
\acmYear{2024} 
\setcopyright{acmlicensed}\acmConference[CHIIR '24]{Proceedings of the 2024 ACM SIGIR Conference on Human Information Interaction and Retrieval}{March 10--14, 2024}{Sheffield, United Kingdom}
\acmBooktitle{Proceedings of the 2024 ACM SIGIR Conference on Human Information Interaction and Retrieval (CHIIR '24), March 10--14, 2024, Sheffield, United Kingdom}
\acmDOI{10.1145/3627508.3638300}
\acmISBN{979-8-4007-0434-5/24/03}

\usepackage{booktabs} 
\usepackage{color}
\usepackage{acronym}
\usepackage{graphicx}

\acrodef{CIS}{Conversational Information Seeking}

\begin{document}

\fancyhead{}
\title{Towards Self-Contained Answers: Entity-Based Answer Rewriting in Conversational Search}

\author{Ivan Sekuli\'c}
\affiliation{%
  \institution{Università della Svizzera italiana}
  \city{Lugano}
  \country{Switzerland}
}
\email{ivan.sekulic@usi.ch}

\author{Krisztian Balog}
\affiliation{%
  \institution{University of Stavanger}
  \city{Stavanger}
  \country{Norway}
}
\email{krisztian.balog@uis.no}

\author{Fabio Crestani}
\affiliation{%
  \institution{Università della Svizzera italiana}
  \city{Lugano}
  \country{Switzerland}
}
\email{fabio.crestani@usi.ch}

\begin{abstract}


\ac{CIS} is an emerging paradigm for knowledge acquisition and exploratory search.
Traditional web search interfaces enable easy exploration of entities, but this is limited in conversational settings due to the limited-bandwidth interface.
This paper explore ways to rewrite answers 
in \ac{CIS}, so that users can understand them without having to resort to external services or sources. 
Specifically, we focus on salient entities---entities that are central to understanding the answer.
As our first contribution, we create a dataset of conversations annotated with entities for saliency.  Our analysis of the collected data reveals that the majority of answers contain salient entities. 
As our second contribution, we propose two answer rewriting strategies aimed at improving the overall user experience in \ac{CIS}.  One approach expands answers with inline definitions of salient entities, making the answer self-contained.  The other approach complements answers with follow-up  questions, offering users the possibility to learn more about specific entities.
Results of a crowdsourcing-based study indicate that rewritten answers are clearly preferred over the original ones. We also find that inline definitions tend to be favored over follow-up questions, but this choice is highly subjective, thereby providing a promising future direction for personalization. 
\end{abstract}



\maketitle

\section{Introduction}


\begin{figure}
    \centering
    \includegraphics[width=\columnwidth]{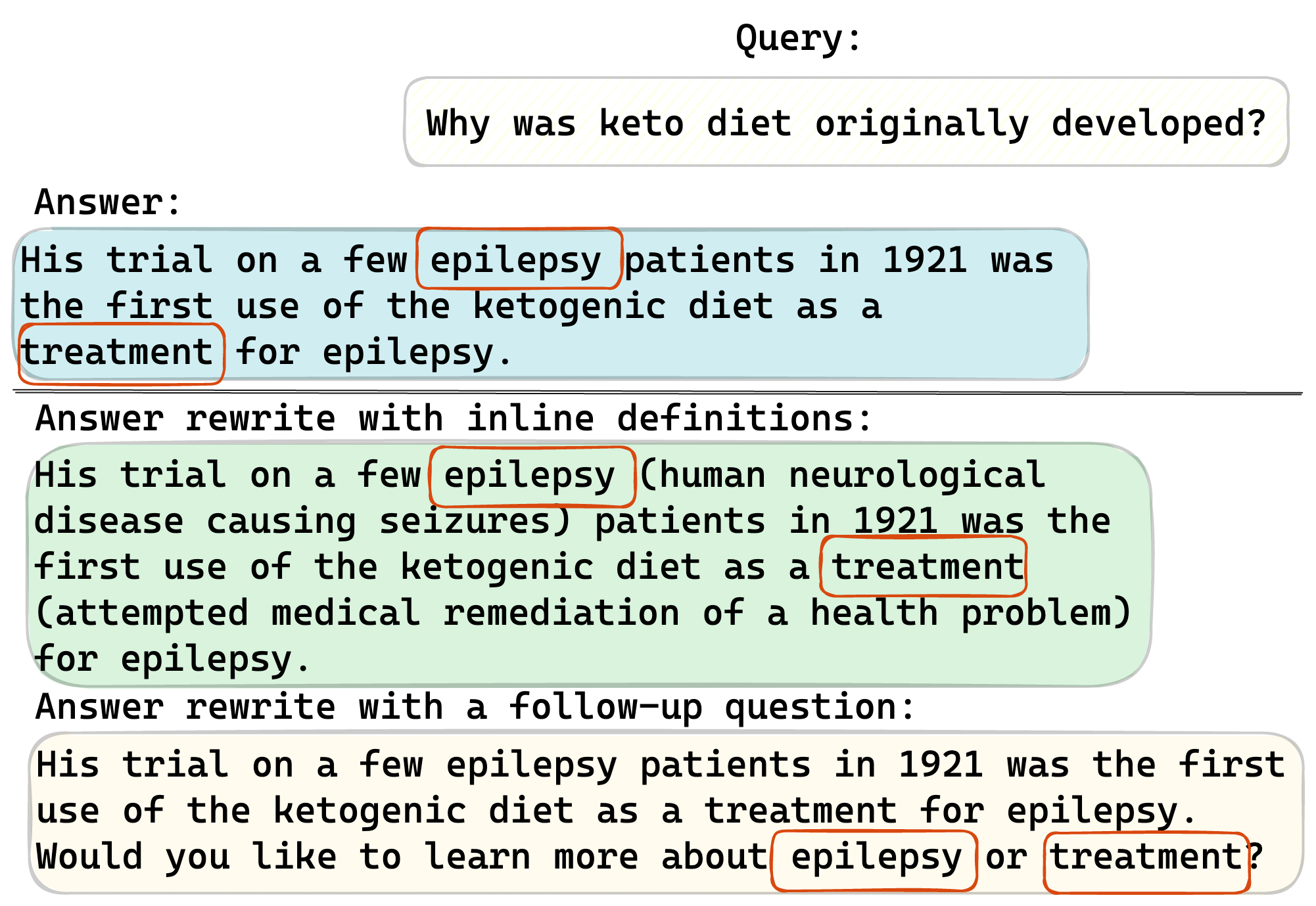}
    \caption{Possible strategies to making sure the user understands the answer in conversational information-seeking.}
    \label{fig:example_rewrite}
\end{figure}

Satisfying users' information needs is the primary goal of any information retrieval system. 
Such search systems are frequently being used for acquiring new knowledge~\cite{Marchionini:2006:Com, Gadiraju:2018:CHIIR}, and enabling effective interaction with them has been the focus of a significant body of research~\cite{Hearst:2009:Book, White:2016:Book}.
With the advent of conversational agents, the landscape of search is changing~\cite{Zamani:2022:FnTIR}, with rapid progress being made in question understanding~\cite{Yu:2020:SIGIR, Vakulenko:2021:ECIR} and result retrieval~\cite{Dalton:2020:arXiv, Yu:2021:SIGIR}.
However, little attention has been paid to supporting users according to their knowledge level~\cite{Ghafourian:2022:ECIR} and ensuring that they can actually understand the answers returned by the system.
While traditional web search offers users the possibility to follow hyperlinks or consult knowledge panels in search engine results pages (SERPs) in order to learn about certain concepts they might be unfamiliar with~\cite{Eickhoff:2014:WSDM}, such opportunity is taken away in conversational information seeking (CIS) due to the limited bandwidth interface.
For example, while the system's generated response might be concise and indeed answer the given question, it might mention concepts that the user is unfamiliar with.
We argue that \ac{CIS} systems offer an unique opportunity to proactively assist an individual---with this work, we aim to make a step in this direction.

Entities are natural units for organizing information and can improve the user experience throughout the search process~\citep{Balog:2018:Book}.
This paper investigates how to make answers more accessible to users in a text-based conversational setting.
The main hypothesis underlying our work is that allowing users to learn more about certain entities mentioned in the answer would lead to an improved user experience. 
However, not all entities are equally important.
Therefore, we utilize the notion of \emph{entity salience} to capture how central a given entity is to understanding the answer returned by the system in response to a question.
Entity salience has been studied in the context of web search, where~\citet{Gamon:2013:CIKM} define it as entities being central and prominent, capturing the aboutness of the Web page.
In this study, we regard entities as anything that could have a Wikipedia page, including named entities, events, and general concepts, borrowing the definition from \citet{Gamon:2013:CIKM}.
While only about 5\% entities are salient in Web pages~\cite{Gamon:2013:CIKM}, 
answers in a conversational setting are short with only a few entities present, therefore yielding a higher ratio of salient entities.
However, not knowing those entities might seriously impair the user's understanding of the answer.
Once the top salient entities are identified, we propose two answer rewriting strategies aimed at helping users to understand the system's response.
One approach rewrites the answer to expand it with inline definitions of salient entities, making the answer self-contained.
The other approach complements the answer with a follow-up question, offering users the possibility to learn more about specific entities.
See Fig.~\ref{fig:example_rewrite} for an illustration.

The first research question we ask is \textbf{(RQ1) What are the characteristics of salient entities in \ac{CIS}?}
To address this question, we conduct an analysis of $360$ answers from well-established conversational Q\&A datasets using crowdsourcing.
Specifically, we extract a number of entities from the answers and ask crowd workers to assess their saliency based on how essential they are to properly understand the answer to the given question.
We find that the majority of the answers contain a number of highly salient entities, providing strong motivation for answer rewriting.
At the same time, our results also suggest that saliency is highly subjective and is likely influenced by the user's background knowledge.
Additionally, we identify categories of salient entities that do not require further definitions as they belong to common sense knowledge or are already explained in the answer.

The second research question we address is \textbf{(RQ2) How to utilize salient entities in answer rewriting for an improved user experience?}
We consider two variants of answer expansion by (1) adding definitions from a knowledge base after the entity mention in parentheses, and (2) inserting human-written descriptions in the text in a natural manner.
Similarly, we study two options for follow-up generation: (1) asking the user directly whether they want definitions of salient entities, and (2) offering an optional follow-up to learn more about specific entities. 
An experimental comparison of these four alternatives using crowdsourcing reveals that users generally prefer some type of answer rewrite over the original answer, with inline definitions being generally favored over answers with follow-up questions.
As part of our experimental protocol, we also ask crowd workers to provide a free-text justification for their choice of answer rewrite preference.
We observe high subjectivity in these responses, with some annotators favoring the original answer for its conciseness, some preferring the one with inline definitions for its comprehensiveness, and others appreciating the conversational nature of answers with follow-up questions.
Overall, our results provide a strong motivation for future research on personalizing answer rewriting, considering both the background knowledge and interaction preferences of users.

Additionally, we explore the potential of using large language models (LLMs) for the entity-based answer rewriting task, given the recent success of LLMs in a wide array of natural language processing and information retrieval tasks \cite{Brown:2020:NIPS, Ouyang:2022:Training, Pereira:2023:ECIR, Gao:2022:ArXiv}. 
Specifically, we experiment with various way of prompting ChatGPT for end-to-end answer rewriting.
Our initial analysis revealed significant shortcomings in terms of knowledge distortion (e.g., rewritten answer contains simpler language, without the original entities the user might want to know about), failure to explain entities, or significantly increasing the answer length, making it unfit for a conversational setting.
Taken together, these issues give rise to concerns regarding the lack of control and faithfulness of the rewritten answers, underscoring the need for more controlled answer rewriting strategies that we are proposing.



Our contributions can be summarized as:
\begin{itemize}
\item We annotate a sample of 360 question-answer pairs to characterize entity saliency in \ac{CIS}.
\item We propose and evaluate two methods for improving the answers given by the search system: rewriting the answer with inline definitions of salient entities and prompting the user with a follow-up question to allowing them to learn more about salient entities.
\item We extensively analyze the feedback on answer rewrite type preference and identify patterns that can help motivate future research.
\item We perform an initial exploration of addressing the same task using a state-of-the-art LLM and provide anecdotal evidence for the need for more controlled generation approaches, thereby solidifying the case for the type of methods this paper is proposing.
\end{itemize}
All resources developed within this paper, including the acquired dataset of salient entities and  crowdsourcing annotations are made available at \url{https://github.com/isekulic/chiir24-answer-rewriting}.

\section{Related work}
\label{sec:related}

We highlight relevant research in the areas of \ac{CIS} and  entity-centric search. As the distinction between conversational search and conversational Q\&A is blurred~\cite{Zamani:2022:FnTIR}, we use \ac{CIS} as an umbrella term.

\subsection{Conversational Information Seeking}
\ac{CIS} has emerged as an increasingly popular method of retrieving information, including the information from the Web \cite{Anand:2020:Dag}.
Several research directions have span from \ac{CIS}, including conversational passage retrieval (e.g., TREC CAsT~\cite{Dalton:2020:arXiv}), conversational Q\&A (e.g., QuAC~\cite{Choi:2018:arXiv}), and mixed-initiative interactions \cite{Zamani:2022:FnTIR}.
Under the mixed-initiative paradigm, the system can at any point proactively take initiative and ask the user clarifying questions or offer suggestions.
While mixed-initiative is a relatively well established concept in the IR community~\cite{Allen:1999:IEEE}, recent advancements in \ac{CIS} systems have demonstrated the effectiveness of asking clarifying questions with a goal of elucidating the underlying user's information need~\cite{Aliannejadi:2019:SIGIR}.
We take advantage of such opportunity and propose to rewrite the answer, offering follow-up to users, as discussed in Section~\ref{sec:rewrite}.
 
\citet{Szpektor:2020:WWW} proposed a dynamic composition-based model for conversational domain exploration (CODEX), which enables users to enrich their knowledge through interactions with the system. 
They highlight several challenges, including maintaining an engaging experience, avoiding repetitions, and choosing the appropriate response length.
While in this work we focus on ensuring user's understanding of answers, some of the points we touch upon are related to the goal of user engagement and not burdening the user with long or repetitive definitions of salient entities.

To the best of our knowledge, response rewriting with the purpose of making sure a user understands the response to their question in \ac{CIS} has not been explored.
However, researchers have studied text rewriting in IR for personalization and text simplification.
While text simplification has been shown to improve readability and understanding in medical~\citep{Leroy:2013:JMIR} and scientific texts~\citep{Ermakova:2022:ECIR}, it is usually done by swapping relatively unfamiliar words with more common alternative words~\citep{Leroy:2013:JMIR} or leveraging large-scale language models for complete rewriting of the text~\citep{Sheang:2021:ICNLG}.
In this setting, a certain degree of information distortion is acceptable, as the text rewritten with such methods might differ from the original due to word substitutions.
On the other hand, we aim to allow the user to learn about a topic of interest, thus retaining the original terminology.



\subsection{Entities}
\citet{Marchionini:2006:Com} categorizes search activities in two broad categories: look-up questions and exploratory search, with the latter requiring carefully curated user interaction~\cite{Camara:2021:CHIIR}.
One of the most notable datasets in the space of web search is the Google Natural Questions dataset~\cite{Kwiatkowski:2019:ACL}, which contains queries from real users with manually evaluated responses.
During their exploratory web search, users often have the possibility to learn about entities of their interest by following hyperlinks or reformulating their query based on newly seen entities~\cite{Eickhoff:2014:WSDM}.
Entity linking and entity-based search are core component in that process \cite{Balog:2018:Book}.
Thus, significant research efforts were put into developing entity linking methods, including entity linking in the Web~\cite{Han:2011:SIGIR}, in free texts~\cite{Piccinno:2014:ws}, and in \ac{CIS}~\cite{Joko:2022:arXiv, Joko:2021:SIGIR}.

While documents may contain a large number of entities, some of them are salient, thus central to modeling the aboutness of a document~\cite{Paranjpe:2009:CIKM}, and others are not.
Moreover, \citet{Gamon:2013:CIKM} find that only about 5\% of the entities in Web pages are salient, while others are often mentioned somewhat sporadically.
These salient entities are crucial for the user to be familiar with, in order to satisfy their information need.
However, in shorter texts that contain fewer entities, this percentage is anticipated to be higher \cite{Wu:2020:NLE}.
Answers in \ac{CIS} are a prime example of such shorter texts.
Yet, research on entity salience in \ac{CIS} is lacking, providing a strong motivation for this work.

Another aspect of entity salience we aim to explore is how important they are for the user's understanding of the texts and readability~\cite{Collins-Thompson:2011:CIKM}.
There is an important distinction to be made between entity salience and entity \emph{relevance} or entity \emph{importance}~\cite{Gamon:2013:CIKM}.
For example, \emph{Joe Biden} is objectively an important entity, however, it can be marginal to the document's topic.
As such, entity relevance is dependent on the user's intent and their underlying information need.
On the other hand, an entity is salient to a document if it is central and important for the overall topical and informational coherency of the document.
Thus, we argue that salient entities are essential to know about for a complete understanding of the provided answers in \ac{CIS}. 
In this work, we explore their prevalence, characteristics, and ways of improving user experience via answer rewriting around identified salient entities.

\section{Understanding Salient Entities in Conversational Information Seeking}
\label{sec:method}

In this section, we define salient entities in \ac{CIS} and present several research questions.
Then, we describe the dataset acquisition process with crowdsourcing.
Finally, we showcase relevant aspects of the created dataset and analyze special cases of salience.

\subsection{Problem statement}
A salient entity captures the aboutness of the text and is thus central to the given document \cite{Paranjpe:2009:CIKM}.
In \ac{CIS}, answers to user's questions are usually short, containing from a single to a few sentences with only a few entities present.
Identifying salient entities in such answers is thus imperative, as they are essential for the user's understanding of the given answer.
In this work, we inspect the prevalence of entity salience in \ac{CIS}.
We define entity saliency on a graded scale of 0 to 2, i.e.,  $s(e_i) \in [0, 2]$, with $e_i$ being the $i$th entity in an answer. A score of 0 corresponds to the entity not being salient at all and 2 to the entity being highly salient.

In this section, we aim to shed light on \textbf{RQ1: What are characteristics of salient entities in \ac{CIS}?}  We break this generic question into a series of more specific subquestions:
\begin{description}
\item[RQ1.a] How prevalent are salient entities in answers in \ac{CIS}?
\item[RQ1.b] How well do users agree on which entities are salient?
\item[RQ1.c] Is there empirical evidence that the notion of entity salience is different in conversational answers than in documents?
\item[RQ1.d] Are there entities that are salient, but do not require explicit definitions?
\end{description}





\subsection{Dataset Acquisition}
\label{sec:expsetup}

In order to model entity salience in \ac{CIS}, we extend QReCC \cite{Anantha:2021:NAACL}---an open-domain conversational question answering dataset containing 14k conversations.
QReCC is curated from three well-established datasets: TREC CAsT 2019 \cite{Dalton:2020:arXiv}, Google Natural Questions (NQ) \cite{Kwiatkowski:2019:ACL}, and QuAC \cite{Choi:2018:arXiv}.
TREC CAsT focuses on conversational passage retrieval, while QuAC resolves around conversational Q\&A over a Wikipedia text.
Contrary, NQ is not conversational in its original form, but has been extended by using its queries as a basis for creating subsequent turns.
Excerpts from QReCC with their original sources and saliency annotations are show in Table \ref{tbl:qrecc}.
All of the conversations in the three datasets have been normalized so that they contain multi-turn interactions with manually resolved utterances and manually checked responses.
This, together with its diversity, makes QReCC appropriate for our work on entity salience in \ac{CIS}.
In this work, we provide a deep analysis of the dataset in terms of entity salience modeling and thus subsample the original QReCC dataset.
We restrict ourselves to the test portion of QReCC, as it contains utterances from all of the three aforementioned datasets.
Additionally, in order to annotate as many conversations as possible within reasonable cost, we restrict ourselves to the conversations up the depth of 3, thereby trading off conversation depth for higher breath coverage.
\begin{table}[]
\caption{Excerpts from QReCC with our crowdsourcing-based annotations of entity salience scores.}
\label{tbl:qrecc}
\resizebox{\columnwidth}{!}{%
\begin{tabular}{@{}llll@{}}
\toprule
Source & Question & Answer & Entity salience score \\ \midrule
CAsT & \begin{tabular}[c]{@{}l@{}}What does it cost to\\ become  a physician's\\ assistant?\end{tabular} & \begin{tabular}[c]{@{}l@{}}Average cost of resident tuition\\ for a 27-month physician\\ assistant program is...\end{tabular} & \begin{tabular}[c]{@{}l@{}}Residency: 1.8\\ Tuition payments: 1.4\end{tabular} \\
\midrule
NQ & \begin{tabular}[c]{@{}l@{}}Why is snow used\\ for igloos?\end{tabular} & \begin{tabular}[c]{@{}l@{}}Snow is used for igloos because\\ the air pockets trapped in it\\ make it an insulator.\end{tabular} & Thermal insulation: 1.5 \\
\midrule
QuAC & \begin{tabular}[c]{@{}l@{}}What was Sigmund Freud\\ and  Wilhelm Fliess'\\ relationship?\end{tabular} & \begin{tabular}[c]{@{}l@{}}(they)\dots saw themselves as \\isolated from the prevailing\\ clinical and theoretical\\ mainstream because of their \\ambitions to develop radical\\ new theories of sexuality.\end{tabular} & \begin{tabular}[c]{@{}l@{}}Human sexuality: 1.0\\ Theory: 0.7\\ Mainstream: 0.4\end{tabular}
\\ \bottomrule
\end{tabular}
}
\end{table}


We employ an established entity linker, WAT~\citep{Piccinno:2014:ws}, to extract entities from the system's responses.
As suggested by the authors, we use a reasonable, slightly precision-oriented confidence threshold of $0.45$ for extracting entities from texts.
We filter out the entities that appear in the question as well, assuming the user asking the question already knows about them.
This procedure results in an entity set $E$, containing several entities extracted from the given answer $A$, that do not appear in the question $Q$.



Now that we have (question, answer, entity\_set) triplets, we employ crowdsourcing to annotate which entities from the entity set can be considered salient.
Given the question and the answer, the task is to annotate the degree to which a given entity is considered essential for understanding the answer.
After an initial analysis of the entities and their importance in understanding the answer, we opted for a graded relevance scale. 
We adopt an annotation scenario where an entity can be either essential, important, or not important.
We draw the similarities of our annotation scheme with well-established graded relevance schemes in IR \cite{Sakai:2021:Book}, where a document can fully satisfy a user's information need, partially, or be irrelevant.
We define the following labels for an entity:
\begin{description}
 \item[Essential.] Knowing about the entity is essential for understanding the answer to the question. 
 It is not possible to comprehend the answer without knowing about (being familiar with) the entity. This label corresponds to a salience score of 2. 
 \item[Important.] Knowing about the entity is important for a deeper and more complete understanding of the answer. 
 However, it is not essential and the user can partially comprehend the answer without knowing about the entity. This label corresponds to a salience score of 1.
 \item[Not important.] The entity is not important for understanding the answer to the question, nor does its knowledge benefit the user's knowledge on the topic. This label corresponds to a salience score of 0.
\end{description}
We use Amazon Mechanical Turk\footnote{https://www.mturk.com} as our annotation platform.
All workers are required to have at least 1,000 approved annotations with a minimum 95\% overall approval rate and be based in the United States, in order to mitigate the potential language barrier for understanding the task.
Each (question, answer, entity) triplet is annotated by five different workers.
To insure high quality annotations, we manually curate a test set of (question, answer, entity) triplets that the workers need to annotate correctly in order for their annotations to count towards the final dataset. 
The size of the test set is 25\% of the final dataset size.
Additionally, we track workers' mouse clicks and discard annotations that are done recklessly and quickly.
Workers take on average $8.1 \pm 12.7$ seconds per (question, answer, entity) triplet.
To ensure ethical use of crowd workforce, we provide an appropriate compensation of 0.20\$ for 5 annotated entities, resulting in an average of 18\$/h, which is over 250\% of the minimum wage in the USA.


\subsection{Analysis}
\label{sec:analysis}
In this section, we answer our research questions through an extensive analysis of the acquired dataset on entity salience in \ac{CIS}.

\subsubsection{Presence of Salient Entities.}
\begin{figure}
    \centering
    \includegraphics[width=0.8\columnwidth]{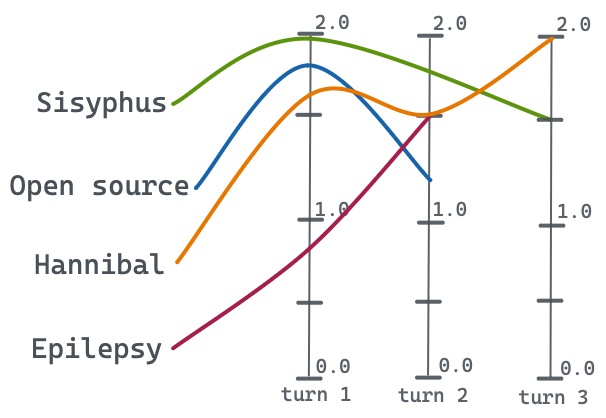}
    \caption{Four examples of changes in salience score through conversational turns.}
    \label{fig:changes}
\end{figure}

\begin{table*}[]
\caption{Examples of special cases of essential entities that do not necessarily require further definitions. Last column indicates the prevalence of such entity types in the expertly-annotated  subsampled set of 122 entities across 37 QA pairs.}
\resizebox{\textwidth}{!}{%
\begin{tabular}{llllr}
\toprule
Special case          & Question                                      & Answer                                                                                                                                       & Entity   & \% in subset   \\
\midrule
Common sense          & \begin{tabular}[c]{@{}l@{}}Who is Sigmund Freud's \\ friend Wilhelm Fliess?\end{tabular} & \begin{tabular}[c]{@{}l@{}}During this formative period of his work...his friend \\ Wilhelm Fliess, a Berlin-based ear, nose, and throat specialist.\end{tabular}                          & Human nose & $25\%$  \\
Location/NE           & \begin{tabular}[c]{@{}l@{}}Did Hansie Cronje \\make any debuts? \end{tabular}           & \begin{tabular}[c]{@{}l@{}}Hansie Cronje made his first-class debut for Orange \\ Free State...at Johannesburg.\end{tabular}                                                               & Johannesburg  & $12\%$\\
Already defined     & \begin{tabular}[c]{@{}l@{}}What are some advantages\\ of using Linux?\end{tabular}      & \begin{tabular}[c]{@{}l@{}}One of the main advantages of Linux is that it is an \\ open source operating system, i.e., its source code \\ is easily available for everyone...\end{tabular} & open source & $2\%$ \\
Entity is the answer & \begin{tabular}[c]{@{}l@{}}Who has to push \\the rock up the hill?\end{tabular}         & \begin{tabular}[c]{@{}l@{}}Sisyphus, king of Ephyra, was punished \\ to roll an immense boulder up a hill... \end{tabular}         & Sisyphus   & $4\%$ \\ \bottomrule
\end{tabular}%
}
\label{tab:essential_concepts_cases}
\end{table*}

In order to focus on answers with a certain level of complexity, we selected answers with at least 2 entities present (as extracted by WAT). This resulted in annotation of 120 QA pairs, containing more than 400 entities.
Each (question, answer, entity) pair was assessed by five different workers, resulting in a total of over 2,000 annotations.
In the annotated dataset, there are on average $5.06 \pm 2.63$ entities present in the answers. 
The average salience of those entities, as assessed by crowd workers, is $1.24 \pm 0.33$ (40\% annotated with salience of 2, 53\% with 1, and 7\% with 0).
In response to \textbf{RQ1.a}, this means that there are more salient entities than non-salient ones in \ac{CIS} answers.
This finding is further confirmed by averaging the saliency scores for each entity and computing the portion of salient ones (e.g., average saliency score > $1.5$) over the total number of entities in the answer.
This ratio is $0.63 \pm 0.28$, meaning that on average 63\% of all entities in \ac{CIS} can be considered salient entities.

Moreover, we analyze salience throughout the conversation.
Figure \ref{fig:changes} shows examples of the development of an entity salience through three turns of the conversation.
The entity \emph{epilepsy} is mentioned sporadically in the answer at turn 1, but becomes considerably more salient in the subsequent turn.
Overall, we observe an average change of saliency score between two consecutive turns of $0.36 \pm 0.21$, suggesting that an entity might become more or less essential as the  focus of the conversation changes.
Entities might be sporadically mentioned in earlier turns of the conversation, but with users' further queries they can become central to the topic of the conversation.

\subsubsection{Subjectivity in Assessing Entity Salience}
To answer \textbf{RQ1.b}, we compute Fleiss' $\kappa$~\cite{Fleiss:1971:Book} to measure subjectivity of the annotators assessing the degree of saliency of an entity, i.e., how essential is the entity for a complete understanding the answer.
The computed $\kappa$ is $0.16$, suggesting weak inter-annotator agreement and high subjectivity for the task~\cite{Viera:2005:Fammed}.
We additionally compute Spearman's rank correlation coefficient $\rho$ between all pairs of workers that annotated a specific QA pair.
With this step, we try to assess potential subjectivity level that is due to different perception of scale of essential/important/unimportant entities.
For example, two workers might agree on which of the entities is more salient, while their perception of the saliency scale differs slightly.
The average Spearman's $\rho$ is 0.45, which suggests a fair agreement and thus a certain level of skewed score subjectivity, which is different from weak agreement measured by $\kappa$.
Overall, we conclude that the task of assessing which entities are essential for answer understanding is highly subjective.
The subjectiveness may come from different user background knowledge, their perception of salience, but also from personalities.
However, having labels collected from five different annotators allows for a robust assessment of entity salience.
The data suggests that there is a lot of potential for dealing with personal preferences and subjectivity when estimating entity salience.

\subsubsection{Entity Salience in Documents vs. in \ac{CIS}.}
We hypothesized that the notion of entity salience is different in \ac{CIS} than in Web documents.
To assess this hypothesis in the light of \textbf{RQ1.c}, we compute the entity salience score using a state-of-the-art model for salience prediction in documents, SWAT~\cite{Ponza:2019:CI}.
For each QA pair, we compute Spearman's $\rho$ over the entities ranked by salience score from the dataset and the entities ranked by salience score as predicted by SWAT. 
The computed $\rho$ averages to 0.25, indicating low to moderate correlation.
This suggest that document-level salience prediction methods are not entirely fit for the task of entity salience identification in \ac{CIS}.
Furthermore, the prevalence of salient entities is significantly higher in \ac{CIS} answers (63\%), as opposed to Web documents (5\%), as reported by \citet{Gamon:2013:CIKM}.


\subsubsection{Special Cases of Salient Entities}
Another important finding of the analysis is the case that although most of the answers contain salient entities, which require user's familiarity to comprehend the answer, not all such entities necessarily require definitions.
To answer \textbf{RQ1.d}, we take a random subsample of more than a hundred entities from the crowd-annotated answers for analysis with a goal of finding potential patterns.
We then perform expert annotation (done by one of the authors of the paper) by carefully inspecting entities in the context of a conversation and note whether they would potentially require explicit definitions or not.
In our analysis, several special cases of entities arose, which might not require further steps to be taken by the \ac{CIS} system, even if deemed salient.

Table \ref{tab:essential_concepts_cases} presents the described cases, with an example and their prevalence, as indicated by the percentage of such entities subsampled set.
We estimate that around $40\%$ of the entities belong to one of the special cases and potentially do not require definitions, with the biggest category being common-sense knowledge entities.

\section{Answer Rewriting}
\label{sec:rewrite}

\begin{figure}
    \centering
    \includegraphics[width=\columnwidth]{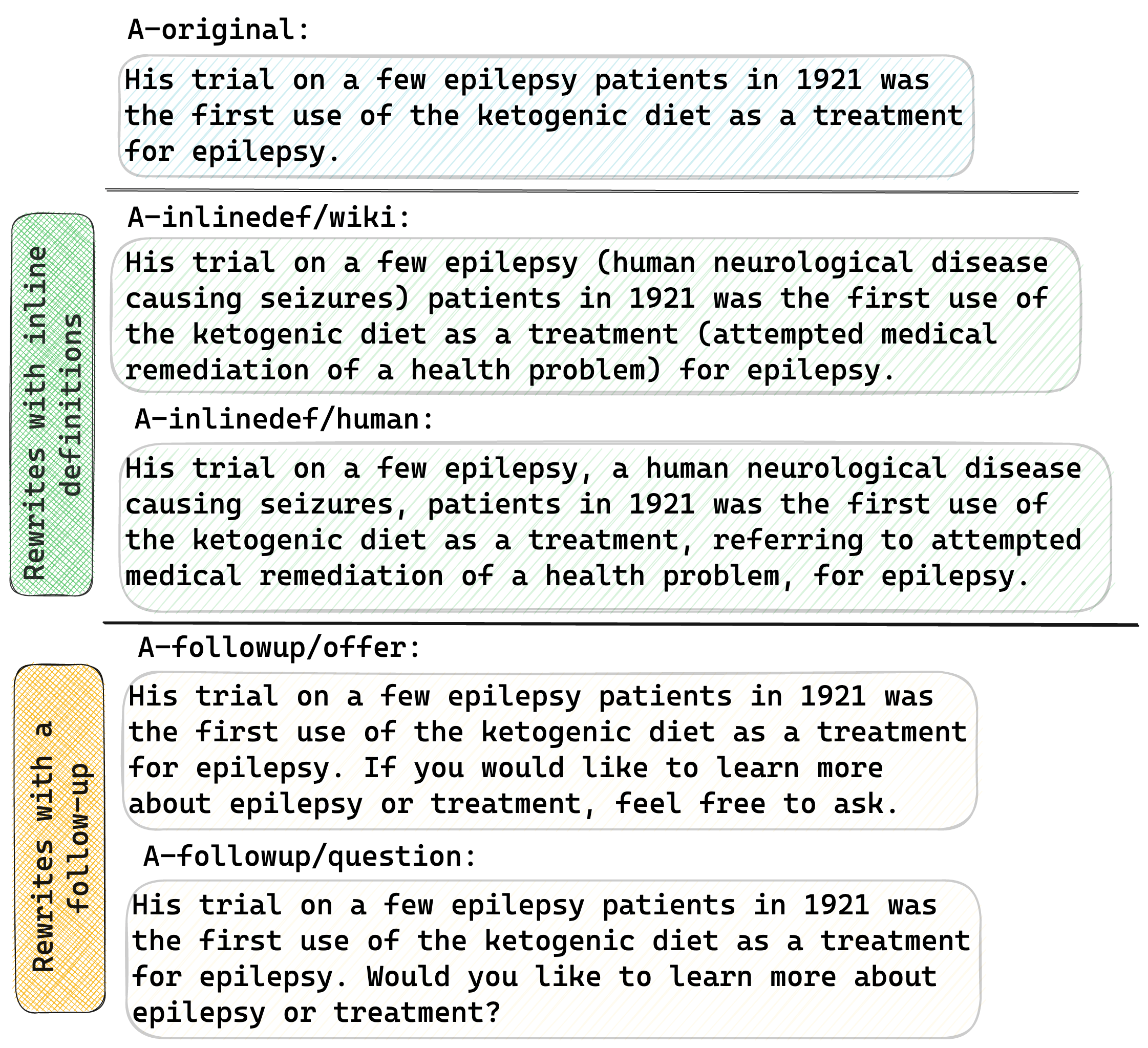}
    \caption{Answer rewriting strategies.}
    \label{fig:rewrites}
\end{figure}

We have established that salient entities occur frequently in answers to \ac{CIS} questions.
In this section, we aim at rewriting the answers containing salient entities with the goal to aid users' understanding.
To this end, we propose two answer rewriting strategies, depicted in Figure \ref{fig:rewrites}.
The first strategy aims to rewrite the original answer $A$ by inserting inline definitions of the identified salient entities, thus making the answer self-contained.
The second strategy makes use of the mixed-initiative \ac{CIS} paradigm and offers the user to learn more about any of the identified salient entities. 
Figure \ref{fig:rewrites} showcases all rewrite types, further explained in the following sections.

\subsection{Inline Entity Descriptions}
Our first strategy towards ensuring the complete understanding of the answer is based on including the explanations of the identified salient entities in the answer itself. 
Formally, we rewrite the original answer $A$ by providing inline definitions $d_i$ for each of the salient entities $e_i$, resulting in the answer rewrite \emph{A-inlinedef}. 
The answer \emph{A-inlinedef} is thus self-contained, as all of the salient entities are explicitly described.
One of the challenges here is to keep the explanations reasonably short and adequate for a conversational setting, as explaining the answer with long definitions would result in a significantly longer answer than the original, thereby overwhelming the user. 
Thus, we experiment with two alternatives for providing inline definitions.

\subsubsection{Wikibase Entity Descriptions.}
We utilize a knowledge base to extract definitions of salient entities.
Specifically, we consult Wikibase\footnote{https://www.mediawiki.org/wiki/Wikibase} to retrieve the entry of given entity $e_i$ and get its definition $d_i$.
To construct the final rewritten answer \emph{A-inlinedef/wiki}, we insert $d_i$ in parentheses immediately after the first mention of $e_i$ ($\dots e_1 \dots e_2 \dots e_3 \dots \Rightarrow \dots e_1 (d_1) \dots e_2 (d_2) \dots e_3 (d_3) \dots$ ).

\subsubsection{Manually Curated Entity Description.}
We hypothesize that entity descriptions inserted into parentheses might appear ineloquent and unnatural for a conversational setting.
Therefore, we manually go through the entity descriptions in \emph{A-inlinedef/wiki} answers and rewrite them to sound more natural.  
The process of manual rewriting involves, among others, avoiding highly technical or too verbose definitions.
Moreover, we insert $d_i$ after $e_i$ such that the definition is blended in the text more naturally, i.e., using commas ($\dots e_1 \dots e_2 \dots e_3 \dots \Rightarrow \dots e_1, d_1, \dots e_2, d_2, \dots e_3, d_3, \dots$ ).
This text simplification task could potentially be carried out by a pretrained large-scale language model.
However, in our initial experiments with T5 \cite{Raffel:2020:JMLR}, we observed several inaccuracies.
As our objective is to measure the usefulness of answer rewrites to users, we opted for human curation to ensure that the findings of this study are not impacted by the imperfections of  automatic rewrites.
The answer rewritten with this method is referred to as \emph{A-inlinedef/human}.


\subsection{Mixed-initiative Follow-up Prompt}
Under the mixed-initiative paradigm in \ac{CIS}, the system can at any point take initiative and prompt the user with various elicitation, clarification, or other questions \cite{Allen:1999:IEEE, Zamani:2022:FnTIR}.
As one of the potential limitations of the previously described approach is overwhelming the user with potentially unnecessary entity definitions, we instead ask the user whether they require the explanations of salient entities or not.
To this end, we experiment with two different follow-up prompts, described below. 

\subsubsection{Follow-up Question.}
The first type of follow-up we propose is a direct question, aimed at asking whether the user is familiar with the salient entities identified in the answer.
To construct a direct clarifying question, we construct a new answer \emph{A-followup/question} by expanding the original answer $A$ with a question ``Do you want to learn more about $e_1$, $e_2$, or $e_i$?'', where $e_i$ is in the top N most salient entities identified.

\subsubsection{Follow-up Offer.}
Similarly, an offered follow-up prompt  (\emph{A-followup/offer}) is designed by expanding the original answer with ``If you wish to learn more about $e_1$, $e_2$, or $e_i$, feel free to ask.''

We hypothesize that this strategy offers several benefits over the inline explanation rewrites.
First, the user can chose whether they want to learn about the identified salient entities or they are comfortable with moving on with the conversation (they either know enough about the entities or do not care).
We note that phrasing the follow-up prompt as a direct question, i.e., ``Do you want to learn more about \textit{entity}?'' would require the direct answer from the user, potentially disrupting the conversation flow.
Instead, our proposed construction of the prompt simply offers the user a possibility for expansion, enabling them to ignore it if they are not interested in learning about the proposed entities.
Second, we can learn about the user's background knowledge by them choosing or not choosing to learn about the salient entities, leading to a potential for personalization of subsequent answers.
Third, we encourage engagement with the user by providing potential topics to converse about.
While these assumptions intuitively make sense, we formulate specific research questions to assess them empirically.

\subsection{Evaluation of Answer Rewrite Strategies}
\label{sec:eval_types}
In this section, we describe the human-based evaluation procedure for comparing the original answer with the rewritten answers.

\subsubsection{Research Questions.}
The main research question we aim to answer is \textbf{RQ2: How to utilize salient entities for answer rewriting for an improved user experience?}
We also aim to explore what type of rewritten answers users prefer and what methods work the best for generating such rewrites.
Thus, we extend our main research question to four more specific questions:
\begin{description}
    \item [RQ2.a:] Do users prefer the rewritten questions over the original ones?
    \item[RQ2.b:] Which of the two answer rewrite strategies (\emph{A-inlinedef} or \emph{A-followup}) is preferred?
    \item [RQ2.c:] Is there a preferred way of explaining the salient entities inline (\emph{A-inlinedef/wiki} or \emph{A-inlinedef/human})?
    \item [RQ2.d:] Is there a preferred way of offering follow up to the user (\emph{A-followup/question} or \emph{A-followup/offer})? 
    \item [RQ2.e:] How does the number of salient entities considered in the rewrite (top 1, 2, or 3) affect user preferences? 
\end{description}

\subsubsection{Experiment Design.}
We design the evaluation study as a multiple choice inquiry and ask crowd workers to provide their assessments.
Given an initial question, the workers need to assume the role of a user and select the answer that they would prefer in an interaction with a conversational assistant.
The given options are threefold: an original answer, a rewritten answer with inline explanations, and a rewritten answer with a follow-up prompt in the end.
Furthermore, to answer research questions RQ2.c and RQ2.d, we vary the methods for inline explanations, as well as the types of questions for the prompt-based rewrite. 
Note that crowd workers are not aware of those changes and they always have the three mentioned options, without knowing how the rewrites are generated.
To ensure consistency, we generate rewrites on the same pool of QA pairs, thus controlling the potential impact of different topics on the rewrite preference. 
Each question and three answer options, corresponding to original answer, an answer with inline definitions, and an answer with follow-up, is annotated by three different crowd workers.



We ensure the quality and consistency of the annotations by selecting high-quality workers, as described in Section~\ref{sec:expsetup}.
Moreover, we randomize the order of \emph{A-original}, \emph{A-inlinedef}, and \emph{A-followup} to reduce any potential position bias.
In order to gain further insights into the underlying rationales, we ask annotators to provide a brief explanation of on why they chose the answer they did.
We analyze the provided reasons in depth in the next section.
To additionally ensure high quality annotations, we manually inspect all of them, rejecting crowd workers who carelessly provided nonsensical reasons (e.g., ``first one,'' ``best text,'' or simply copy-pasted parts of the answers), and blocking them from further participation in the study.
In total, we acquire more than 600 assessments on rewrite type preference with justifications for the choice.
\subsection{Results}
\label{sec:results}

In this section, we present the results of the crowdsourcing study on answer rewrite type preference and analyze them in the light of the aforementioned research questions.
\vspace{-0.5mm}
\subsubsection{Original or Rewritten Answer Preference}
\begin{table}[]
\caption{Answer rewrite preference assessed by crowd workers. Workers are shown three rewritten answers (\emph{A-original}, \emph{A-inlindef}, and \emph{A-followup}) in a randomized order. Results are broken down by varying the proposed answer rewrite strategies. p-values are reported using a $\chi^2$ test.}
\label{tbl:results_types}
\resizebox{\columnwidth}{!}{%
\begin{tabular}{r|rr|rr|r}
\toprule
\multicolumn{1}{c}{Original} & \multicolumn{2}{c}{\emph{A-inlinedef}} & \multicolumn{2}{c}{\emph{A-followup}} & \multicolumn{1}{c}{p-value} \\
\multicolumn{1}{c}{} & \multicolumn{1}{c}{/wiki} & \multicolumn{1}{c}{/human} & \multicolumn{1}{c}{/offer} & \multicolumn{1}{c}{/question} & \multicolumn{1}{c}{} \\
\midrule
60 & 66 & - & 45 & - & 0.13 \\
56 & 53 & - & - & 41 & 0.28 \\
54 & - & 60 & 36 & - & 0.04 \\
52 & - & 71 & - & 27 & $<0.01$ \\
\midrule
222 & \multicolumn{2}{c|}{250} & \multicolumn{2}{c|}{149} & $<0.01$ \\
\bottomrule
\end{tabular}
}%
\end{table}

Table \ref{tbl:results_types} presents the results of the different combinations of answer rewrites, as explained in Section \ref{sec:eval_types}.
To assess whether differences in answer preference are statistically significant, we perform a $\chi^2$ test under the null hypothesis of data being drawn from a uniform probability distribution across the three rewrites (i.e., each row of the table).
In response to \textbf{RQ2.a}, we observe a preference for one of the answer rewrites, over the original answer (222 for original vs 399 for rewrites, p-value $< 0.05$).
These results suggest that there is a large potential for improving the user experience through answer rewriting.
Moreover, the findings suggest a promising direction for further research on answer rewriting in \ac{CIS} systems, both by providing further inline explanations of certain entities and by offering follow-up clarifications.

\vspace{-0.5mm}
\subsubsection{Rewrite Type Preference}
Regarding \textbf{RQ2.b}, we observe a preference for answers with inline explanations (\emph{A-inlinedef}) over the answers with a follow up (\emph{A-followup}).
Moreover, as indicated in Table \ref{tbl:results_types}, this preference is prevalent across all combinations of rewrite subtypes.
Although not all combinations in Table \ref{tbl:results_types} yield statistically significant differences, the overall trend is prevalent across all of the experiments.
This suggests that making the answer self-contained by providing inline entity explanations is more desirable than offering the user to clarify these entities.
Contrary to our hypothesis, longer answers obtained by inserting entity descriptions do not seem to overwhelm the majority of the users.
However, subjectivity is still important in this scenario, as some users indeed find \emph{A-inlinedef} to be too cluttered, as discussed in the next section.

\vspace{-0.5mm}
\subsubsection{Rewrite Subtype Preference}
To address research questions \textbf{RQ2.c} and \textbf{RQ2.d}, we aggregate the results of different subtypes of answer rewriting.
Experiments indicate humanly-curated answer to be slightly more preferred over the Wikbase definitions in parentheses (131 vs. 119), suggesting that more natural rewrites could better help the user understand the answers.
This finding is a motivation for the development of answer rewriting methods aimed at defining entities in a more natural manner, compared to entity definitions being inserted into parentheses.

Similarly, \emph{A-followup/offer} is slightly more preferred than \emph{A-followup/question} (81 vs. 68).
We hypothesize that a prompt that could be ignored, as opposed to a direct question, would benefit the overall user experience.
While both strategies are equally effective in providing the user with desired information, \emph{A-followup/offer} might not impair the flow of the conversation, as it can be ignored if the user does not desire to learn more about proposed entities.



\vspace{-0.5mm}
\subsubsection{Result Preference by Top N Entities}

Regarding \textbf{RQ2.e}, we report the results on answer preference across top $N$ most salient entities rewritten in Table \ref{tbl:topN_results}.
Specifically, we construct the experiment such that the same original answer is rewritten three times, each time with $N$ salient entities taken into account, with $N \in [1,2,3]$.
Results suggest that the higher the $N$, i.e., the more entities are defined in the answer, the stronger the user's preference for \emph{A-inlinedef}.
We hypothesize that such answers provide a more complete response to the given question, thus not requiring further explorations of the topic through clarifying prompts.

\begin{table}[]
\caption{Results on answer rewrite preference by top N salient entities rewritten. The p-value is computed by $\chi^2$ test.}
\label{tbl:topN_results}
\begin{tabular}{llllr}
\toprule
 & $A$ & \emph{A-inlinedef} & \emph{A-followup} & p-value \\
 \midrule
Top 1 & 80 & 84 & 65 & $0.27$ \\
Top 2 & 87 & 83 & 42 & $<0.01$ \\
Top 3 & 55 & 83 & 42 & $<0.01$ \\
Total & 222 & 250 & 149 & $<0.01$ \\
\bottomrule
\end{tabular}
\end{table}

\subsection{Analysis}
\begin{table*}[]
\small
\caption{Reasons for choosing one answer over another, grouped together by observed patterns. The last column presents the prevalence of the pattern in the manually analyzed portion of the dataset.}
\centering
\label{tbl:reasons_final}
\resizebox{\textwidth}{!}{%
\begin{tabular}{llll}
\toprule
Reason for choice & Preference & Pattern & Prev. \\
\midrule
\begin{tabular}[c]{@{}l@{}} $*$ ``The terms seem to be very specialized in terms of the question  and need at least some elucidation to understand''\end{tabular} & \emph{A-inlinedef} & \multirow{2}{*}{\begin{tabular}[c]{@{}l@{}} Better \\explanation\end{tabular}} & \multirow{2}{*}{28\%} \\
$*$ ``not everyone knows how the body functions.'' & \emph{A-inlinedef} &  &  \\
\midrule
\begin{tabular}[c]{@{}l@{}}$*$ ``The most concise answer. I don't think the recipient would \\ want any more particular information especially about Africa''\end{tabular} & \emph{A-original} & \multirow{2}{*}{Concise} & \multirow{2}{*}{28\%} \\
\begin{tabular}[c]{@{}l@{}}$*$ ``it's the most complete answer that doesn't add unnecessary  stuff inside brackets of brackets.''\end{tabular} & \emph{A-original} &  &  \\
\midrule
$*$ ``the other answers have text that shouldn't be there or is too wordy.'' & \emph{A-original} & \multirow{2}{*}{No clutter} & \multirow{2}{*}{15\%} \\
\begin{tabular}[c]{@{}l@{}}$*$ ``I prefer this one because it doesn't have a question on the end, and \\ because it is the most clear and direct, w/o any parentheticals.''\end{tabular} & \emph{A-original} &  &  \\
\midrule
$*$ ``invites you to ask for more information about it'' & \emph{A-followup} & \multirow{2}{*}{Learn more} & \multirow{2}{*}{14\%} \\
$*$ ``I choose 1, because I want to learn more about solar energy.'' & \emph{A-followup} &  &  \\
\midrule
\begin{tabular}[c]{@{}l@{}}$*$ ``I prefer the first one because it is the most concise answer. One doesn't need to \\ be  told they ask more questions as in answer 2, and answer 3 rambles a bit.''\end{tabular} & \emph{A-original} & \multirow{2}{*}{Other} & \multirow{2}{*}{10\%} \\
\begin{tabular}[c]{@{}l@{}}$*$ ``the topic is very familiar for me and i have some knowledge about chemical energy so i choose this option''\end{tabular} & \emph{A-inlinedef} &  &  \\
\midrule
\begin{tabular}[c]{@{}l@{}}$*$ ``This doesn't overexplain the compatibility layer aspect, making me feel overwhelmed, \\ and allows me to ask about it if I want instead''\end{tabular} & \emph{A-followup} & \multirow{2}{*}{Natural} & \multirow{2}{*}{4\%} \\
\begin{tabular}[c]{@{}l@{}}$*$ ``Apart from answering the question, the assistant is more  interactive and \\ continues to ask whether I would like to learn more about the bank of England.''\end{tabular} & \emph{A-followup} &  & 
\\ \bottomrule
\end{tabular}
}
\end{table*}

In order to gain further insight into answer rewrite preferences, we manually analyze  responses from crowd workers.
Recall that workers were asked to justify why they would prefer the answer rewrite they have chosen. 
We perform a qualitative analysis of the provided reasons by identifying re-occurring reasons for workers' choices.
We find five distinct patterns of the provided reasons, presented in Table~\ref{tbl:reasons_final}. 
To estimate the prevalence of each of these, we randomly select 100 responses from the crowd workers and label them using these pattern.
In our analysis, we allow for multiple patterns to be associated with a single reason of choice, as for example worker can pick an answer because it is both \textit{concise} and \textit{natural}.
The patterns and their frequency in the analyzed set are shown in Table~\ref{tbl:reasons_final}.
Overall, we find that:
\begin{itemize}
    \item Users prefer the original answer $A$ mostly because it is short, concise, and does not introduce unnecessary clutter. 
    \item Users prefer the answer with inline explanations \emph{A-inlinedef} when they appreciate additional information and think it provides a well-defined answer.
    \item Users prefer the answer with a follow-up prompt \emph{A-followup} when they think it is the most welcoming of the different answer alternatives, inviting for further conversation, but is not unnecessarily overwhelming with long explanations of entities they perhaps do not require explanation of.
\end{itemize}
\noindent
Although all of the workers provided reasonable justifications for their selection, the inter-annotator agreement, as measured by Fleiss' $\kappa$, is $0.03$, indicating high subjectivity.
From the conducted experiments and observed patters in user's preferences, we find solid evidence that entity-based answer rewriting can lead to an improved conversational user experience. At the same time, we also find that the choice of preferred answer format is highly subjective, which calls for further research on the personalization of such approaches.

\vspace{-2mm}
\subsection{LLM-based Answer Rewriting}

\begin{table*}[]
\small
\caption{Examples of LLM rewrites for answer rewriting with the original answer being: \textit{``The lipid bilayer is a thin polar membrane made of two layers of lipid molecules. These membranes are flat sheets that form a continuous barrier around all cells.''}}
\label{tbl:LLM_examples}
\resizebox{\textwidth}{!}{%
\centering
\begin{tabular}{lll}
\toprule
Prompt & Answer rewrite & Limitation \\
\midrule

\begin{tabular}[c]{@{}l@{}}Rewrite the passage below in a \\simplified language, so it is \\understandable to everyone.\end{tabular}                                                               & \begin{tabular}[c]{@{}l@{}}
The lipid bilayer is a very thin and special kind of membrane made up of two\\ layers of lipid molecules. These membranes are \textit{like flat sheets} that wrap\\ around all cells and create a protective barrier. \end{tabular}                                                                                          & Knowledge distortion   \\
\midrule
\begin{tabular}[c]{@{}l@{}}Rewrite the passage below so it contains\\ explanations of the main concepts\\ so that it is easier to understand.\end{tabular}                                                                     & \begin{tabular}[c]{@{}l@{}}The lipid bilayer is a structure consisting of two layers of lipid \\molecules that form a thin, polar membrane. This membrane serves as a\\ protective barrier encircling every cell in the body. It is composed \\of flat sheets, ensuring a continuous and consistent covering.  \end{tabular}                                                  & No entity explanations \\
\midrule
\begin{tabular}[c]{@{}l@{}}Rewrite the passage below so it is\\ understandable  to wide audiences, by\\ explaining the following  entities: \\"lipid molecules", "polar membrane".\end{tabular} & \begin{tabular}[c]{@{}l@{}}The cells in our body are surrounded by a very thin\dots\\ head is attracted to water (hydrophilic) while the \\ tail is repelled by it (hydrophobic)\dots\\ \textit{(5 sentences)}.\end{tabular} & Answer lengthening    
\\ 
\midrule
\begin{tabular}[c]{@{}l@{}}Rewrite the passage below so it contains\\ brief explanations of  "lipid molecules"\\ and "polar membrane".
\end{tabular}                                                                     & \begin{tabular}[c]{@{}l@{}}
The lipid bilayer, which is made up of lipid molecules, is a thin and polar\\ membrane. Lipid molecules are special types of fats. This polar membrane\\ consists of two layers and forms flat sheets. It acts as a continuous\\ barrier that encloses and protects all cells.
\end{tabular}                                              & -- \\
\bottomrule
\end{tabular}
}%

\end{table*}

While entity definitions are currently either taken from a knowledge base or curated by human intervention, the question naturally arises: Could this task not be tackled in an end-to-end manner by a large language model (LLM)?
Given promising advancements in text simplification, e.g., with T5~\cite{Sheang:2021:ICNLG}, and the broad variety of knowledge and language capabilities demonstrated by ChatGPT, there are reasons to believe that the answer rewriting task could be performed by simply engineering the ``right'' prompt. Below, we present some anecdotal evidence to the contrary, by presenting results obtained with a state-of-the-art LLM, \texttt{gpt-3.5-turbo}~\cite{Brown:2020:NIPS}.

Specifically, we consider two types of prompts: (1) instructing the LLM to rewrite the answer for easier understandability and (2) additionally, including the specific entities that need to be explained.  For both settings, we experimented with both zero- and few-shot prompts. 
Due to space constraints, we only include a few examples in Table~\ref{tbl:LLM_examples} to illustrative the main limitations we identified:
\begin{itemize}
    \item \emph{Failure to identify salient entities}: When entities that require explanations aren't explicitly stated, the LLM can simply reword the answer, without providing any additional information.
    \item \emph{Knowledge distortion}: Certain salient entities are removed from the original answer, causing the loss of information by oversimplifying the text.
    \item \emph{Answer length}: when explicitly stated which entities require inline explanations, the LLM tends to significantly lengthen the original answer (from 1-2 to 5-6 sentences).
    \item \emph{Inconsistency}: Although hallucination is a known issue in LLMs~\cite{Ji:2023:ACMCS}, we also observe inconsistency, i.e., a high degree of variation in answer quality, when generating answers to the same prompt multiple times (controllable with parameters to some degree) and across different examples (not controllable).
\end{itemize}
That said, LLMs can also generate appropriate rewrites, as illustrated by the last example in Table~\ref{tbl:LLM_examples}, which is both concise and natural, while still covering all of the salient entities.
However, it is evident that the salient entities had to be explicitly stated and that prompts need to be carefully engineered for the desired outcome.

The main take home message of our study is that entity-based answer rewriting can improve the user experience, but to unlock its full potential, the identification of salient entities as well as the preferred form of answer rewrite need to be addressed in a personalized manner. These parts require future research. When it comes to the actual generation of the rewritten answer, there is a large potential for utilizing LLMs, provided that they are prompted with the specific entities and the desired format of rewrite.



\vspace{-2mm}
\subsection{Discussion}
\label{sec:results:discussion}
Ours is a novel task in a conversational setting, which makes evaluation inherently challenging.
In this section, we reflect on some of the design decisions, acknowledge limitations, and highlight possible future research directions, including potentially revisiting some of the design choices.

\vspace{-0.5mm}
\subsubsection*{Impact of Rewriting on Answer Length}
Rewriting answers in \ac{CIS} by inserting inline definitions of salient entities lengthens the original answer.
    As observed in our experiments, up to three entity definitions do not seem to hurt the answer rewrite, as such rewrite was often chosen by the crowd workers.
    However, in case the answer becomes too long due to a large number of salient entities, the amount of them we provide definitions for can be reduced by taking only the top N entities, as ordered by the salience scores.

\vspace{-0.5mm}
\subsubsection*{Text- vs. Voice-based CIS}
 We hypothesize that results on answer rewrite preference might differ in a voice-only setting, as the user is not able to skim through potentially unnecessary parts of the answer.
As such, preference for inline definitions might not be so prevalent, as users could not simply skim through the text and would in fact need to listen to the extended answers.
We aim to explore the aforementioned questions in further research.

\vspace{-0.5mm}
\subsubsection*{More Realistic Conversational Setting} Design-wise, we compare answer rewrites turn by turn, rather than evaluating the whole conversations.
This is often the case in crowdsourcing-based studies due to the limited availability of users, although recent research points out the benefits of multi-turn dialogue evaluation \cite{Li:2019:arXiv}.
At the same time, utterances in our study are self-contained and do not necessarily require full conversation history for correct assessments.
Also, we provide an analysis of the salient entity evolution throughout the conversation.
Nevertheless, as part of our future work, we aim to build multiple \ac{CIS} systems based on answer rewrite type (e.g., a system that generates answers with inline explanations of salient entities and a system that offers follow-up prompts) and perform a thorough user study to validate the findings of this work.

\vspace{-0.5mm}
\subsubsection*{Salient Entity Annotation}
While other automated options for extracting salient entities exist, we opted for annotating salience through crowdsourcing with a goal of acquiring high-quality data. 
Nevertheless, despite having multiple controls in place for ensuring quality (from the selection of crowd workers to using test questions), the inter-annotator agreement turned out to be relatively low.
We attribute this to the high subjectivity of the task, as workers' perception of what is ``essential to understand'' might differ, in relation to their personal knowledge and their understanding of what ``essential'' means. We acknowledge the possibility of the annotation task being set up this way to be too open for interpretations, or simply too hard, thus leading to low inter-annotator agreement.  In the future we plan to repeat the annotation process as part of a dedicated study, aiming to untangle what role prerequisite knowledge and subjectivity might play here. 
Nonetheless, we believe the findings of this study on question rewrite strategies and preferences to be sound and useful for the research community.

\vspace{-0.5mm}
\subsubsection*{The Role of Background Knowledge}
High subjectivity of rewrite preferences potentially comes from different backgrounds and personal preferences of different users.
Thus, having information about the background of the user would help the system tailor the rewrite to a specific user, both by selecting only entities which the user needs explanation for and by adjusting the style of the rewrite.
Future work therefore includes personalized answer rewrites.




\vspace{-2mm}
\section{Conclusions}
\label{sec:concl}

In this study, we analyzed the presence of salient entities in conversational information seeking interactions.
We found that most of the answers generated by the search system contain some amount of salient entities, required for the complete comprehension of the answer.
Moreover, with a goal of ensuring that the user understands these answers, we proposed two strategies for answer rewriting.
The first one is based on providing inline definitions of salient entities, while the second one explicitly offers the user to learn more about the entities they might be unfamiliar with.
The suggested methods were extensively assessed through human-based evaluation, indicating user preference for answers with inline definitions, over the follow-up prompt-based rewrites.
We hope that these findings provide a strong motivation for further research on entity-based answer rewriting.


\bibliographystyle{ACM-Reference-Format}
\bibliography{chiir2024-rewriting.bib}
\balance

\end{document}